\documentclass[runningheads]{llncs}
\usepackage[numbers,sort]{natbib}
\usepackage{graphicx}
\usepackage{xcolor}
\usepackage{booktabs} 
\usepackage{multirow} 
\usepackage{appendix}
\usepackage{amsmath}
\usepackage{amssymb}
\usepackage{animate}
\usepackage{media9} 
\usepackage{multimedia}

\newcommand{\numberset}[1]{\mathbb{#1}} 
\newcommand{\real}{\numberset{R}}

\begin{document}

\title{3D Spine Shape Estimation from Single 2D DXA}
\author{Emmanuelle Bourigault \and
Amir Jamaludin \and
Andrew Zisserman}
\institute{Visual Geometry Group, Department of Engineering Science,\\University of Oxford, Oxford, UK\\
\email{emmanuelle@robots.ox.ac.uk} }

\maketitle

\begin{abstract}
Scoliosis is traditionally assessed based solely on 2D lateral deviations, but recent studies have also revealed the importance of other imaging planes in understanding the deformation of the spine. Consequently, extracting the spinal geometry in 3D would help quantify these spinal deformations and aid diagnosis. 
In this study, we propose an automated general framework to estimate the {\em 3D }spine shape from {\em 2D} DXA scans. We achieve this by explicitly predicting the sagittal view of the spine from the DXA scan. Using these two orthogonal projections of the spine (coronal in DXA, and sagittal from the prediction), we are able to describe the 3D shape of the spine.
The prediction is learnt from over 30k paired images of DXA and MRI scans. We assess the performance of the method on a held out test set, and achieve high accuracy\footnote{Code is available at 
\url{https://www.robots.ox.ac.uk/~vgg/research/dxa-to-3d}.}

\end{abstract}

\section{Introduction}

The standard procedure to examine the spine for the presence of scoliosis is using antero-posterior (AP) X-rays and measuring the angle between the most tilted vertebrae~\cite{cobb1948outline}. Scoliosis typically affects growing children and proper diagnosis of scoliosis requires multiple follow-up scans. As such, Dual-energy X-ray (DXA) scans, with its lower radiation dose than X-rays is quickly becoming an acceptable alternative~\cite{taylor2013identifying,Jamaludin2018PredictingSI,NgDEXAuse,jamaludin2023predicting}. 
However, this still does not solve the fundamental issue of scoliosis diagnosis; it is essentially a 3D disorder but the focus is on the 2D lateral shift of the spine. 
Imaging this disorder on a Magnetic Resonance Imaging (MRI) is a good alternative, and multiple studies have shown that there is indeed more useful information that can be extracted from MRIs for scoliosis~\cite{Ills2010BreakthroughIT,3DScoliosis,Bourigault23}. MRI however is more expensive and requires more time for one single scan compared to X-ray/DXA. 

In this paper, our objective is to obtain the 3D patient-specific spine shape from a 2D DXA. We explore whether this is even possible given that 3D information is ``lost'' in the projection of a 2D DXA scan. We show that it is possible to learn to infer this ``lost" information by leveraging a paired imaging set of DXA scans with corresponding  MRIs taken at roughly the same time. The MRI provides the 3D geometry of the spine, and a network can be trained to map the 3D spine from the 2D DXA. We achieve this task by regressing coronal and sagittal curve projections. Through these two orthogonal curves, coronal and sagittal, we can recover the 3D geometry of the spine.

\begin{figure}[t]
    \centering
    \includegraphics[width=1\linewidth]{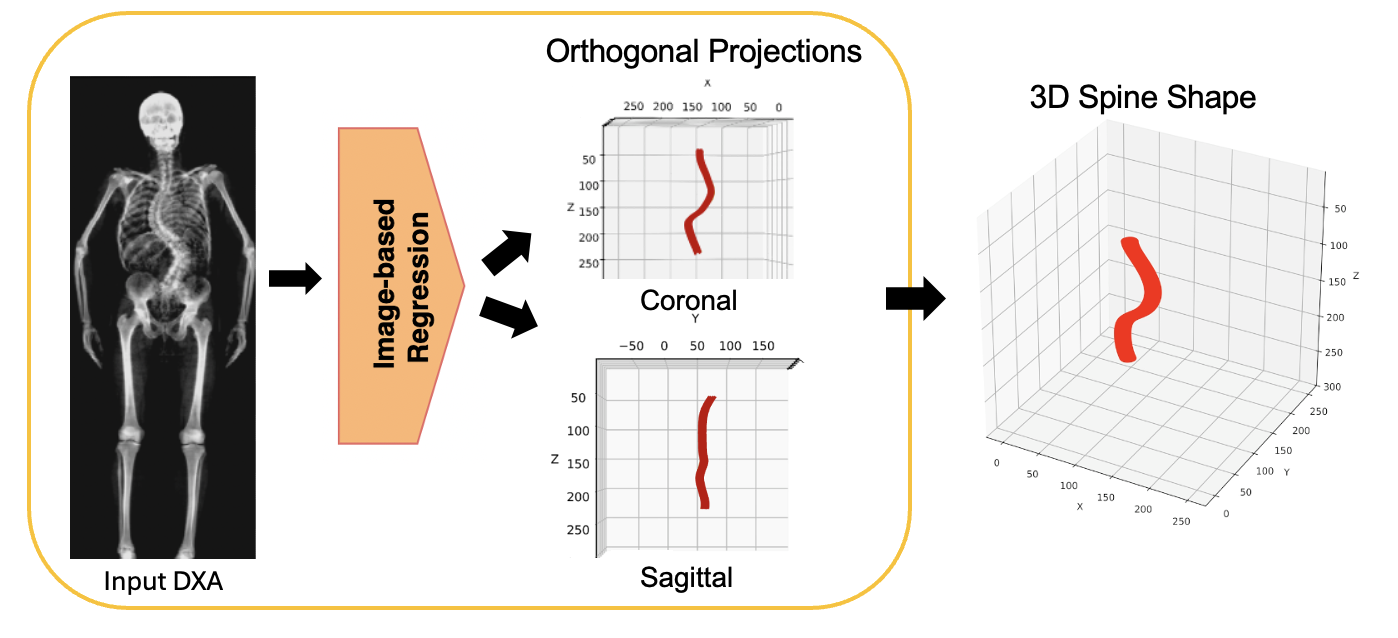} 
    \caption{\textbf{Inference}. Given a DXA scan, the model predicts the coronal and sagittal projections of the 3D spine. Once these two orthogonal views of the spine are obtained, the 3D spine can be reconstructed. A visualisation of the rotating spine is given at the website \url{https://www.robots.ox.ac.uk/~vgg/research/dxa-to-3d}.}
    \label{fig:overview}
\end{figure}

\subsection{Related Work}

To date, the vast majority of scoliosis research has focused on 2D shape analysis of the spine or part of the 3D spine~\cite{taylor2013identifying,Jamaludin16,Jamaludin2018PredictingSI,windsor2020,bourigault2022scoliosis,NgDEXAuse,jamaludin2023predicting}.
Limitations of 2D spine analysis arise particularly in classifying 3D curve shape. Indeed, deviations are not limited to the coronal plane, they include twisting of the spine in multiple directions~\cite{Rockenfeller}, and the importance of other planes, e.g.\  for axial rotations, is well recognized~\cite{Ills2019TheTD}. 
Recently, there has been a growing interest in the community on EOS imaging with simultaneous acquisition of coronal and sagittal views producing 3D spine shape~\cite{Rehm20173DmodelingOT}. This enables, for example, a better evaluation of the effect of brace surgical treatment~\cite{Courvoisier2014,Dubousset2014UseOE}. 

There are several works with similar goal to ours; the most similar work is by~\cite{López2018} which works on DXA scans and predicts the 3D model of the spine using statistical shape models (SSM). Other works use biplanar X-rays either with SSM~\cite{aubert2019toward,benameur20033d,clogenson2015statistical} or contour matching~\cite{zhang20133}. In our work, we directly estimate the 3D spine from a single 2D DXA. 
Our 3D regression involves only a single pass through a feed-forward network.

\section{3D From 2D}

Our method for estimating 3D spine geometry from a single 2D DXA is simple. We essentially learn to regress 2D  curves; first by directly regressing the curve of the spine extracted from the 2D DXA itself (this is a coronal projection of the 3D spine), and then by predicting the sagittal projection of the  3D spine curve.
Given these two orthogonal projections (coronal $x(z)$ and sagittal $y(z)$), the 3D curve can trivially be obtained (as $(x(z), y(z))$). 
Learning the sagittal projection is only feasible through the use of a large-scale public dataset, consisting of {\em paired} whole-body DXA and MRI of the same subjects~\cite{Sudlow2015UKBA}, where the 3D spine curve can be extracted from the MRI. The challenges of this problem are: (i) alignment of the paired DXA and MRIs,  discussed in Section~\ref{sec:alignment}, and (ii) how to directly regress 2D curves from the DXA scan, discussed in Section~\ref{sec:regression}.
\begin{figure}[t]
    \centering
    \resizebox{1\linewidth}{!}{
    \tabcolsep=0.05cm
        \begin{tabular}{cccccc}
DXA & MRI & $1^{st}$ Stage & Spines Not Aligned & $2^{nd}$ Stage \\ 
            \includegraphics[trim= 0.0cm 1.7cm 0.0cm 0cm,clip, width=0.16\linewidth]{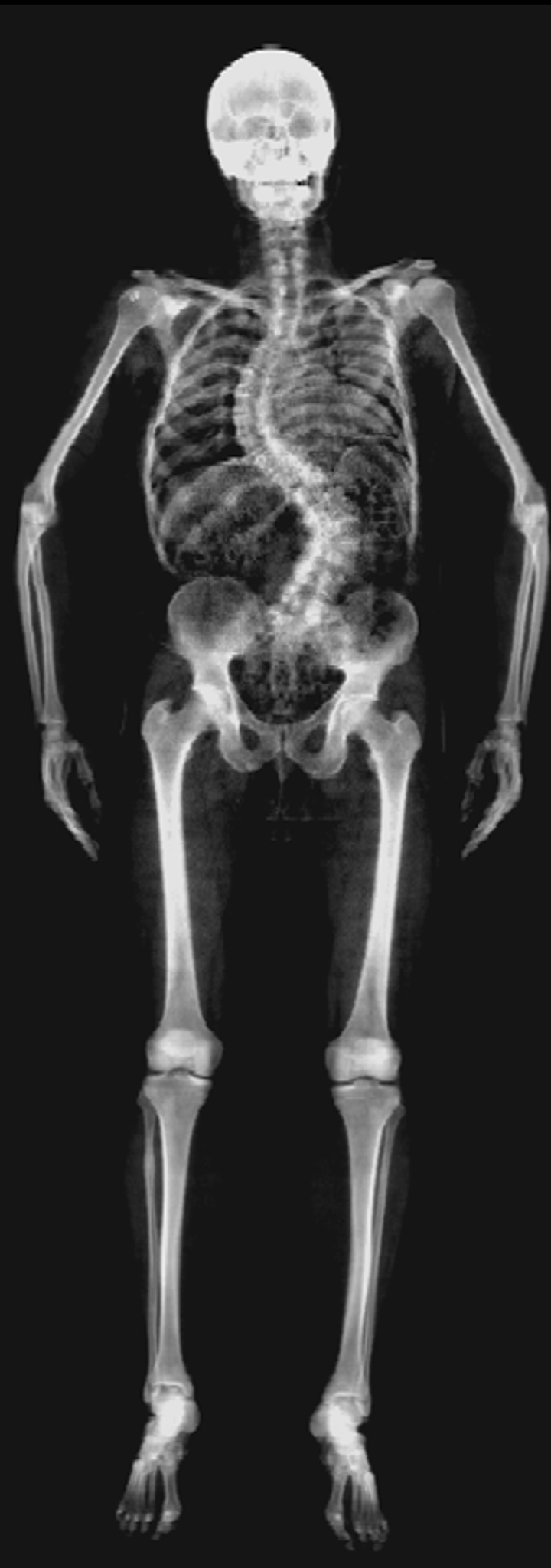}& 
            \includegraphics[trim= 0.0cm 1.7cm 0.0cm 0cm,clip, width=0.16\linewidth]{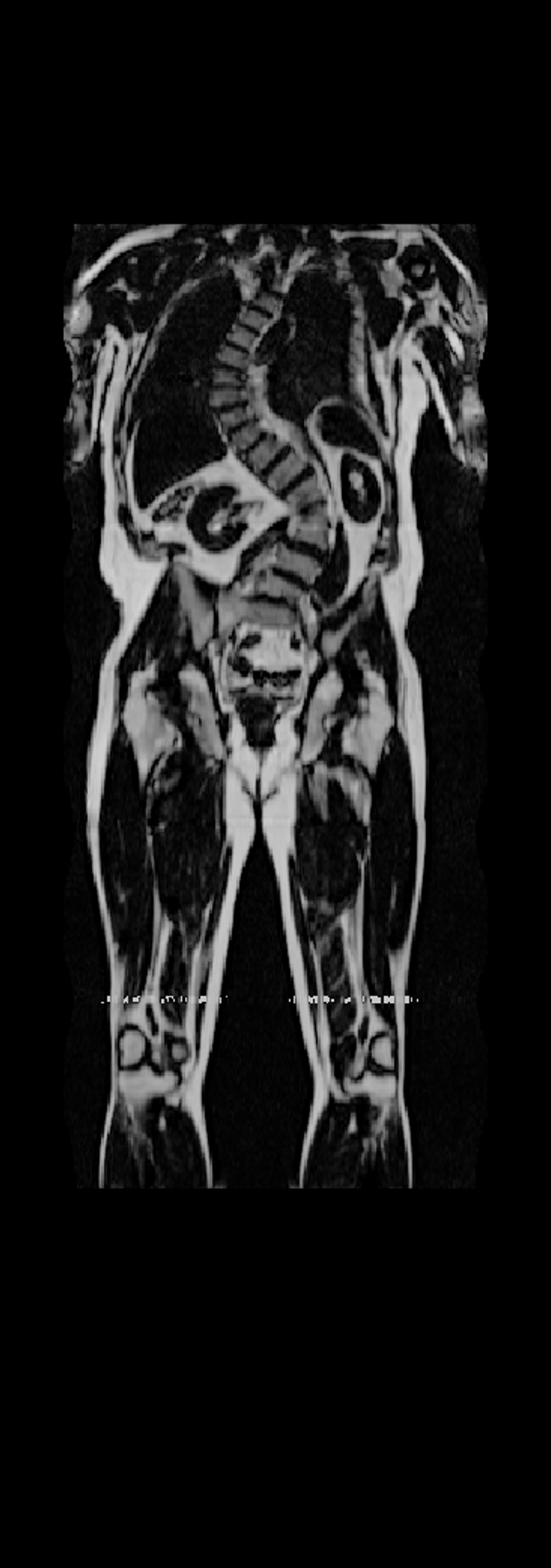}& 
            \includegraphics[trim= 0.0cm 1.7cm 0.0cm 0cm,clip, width=0.16\linewidth]{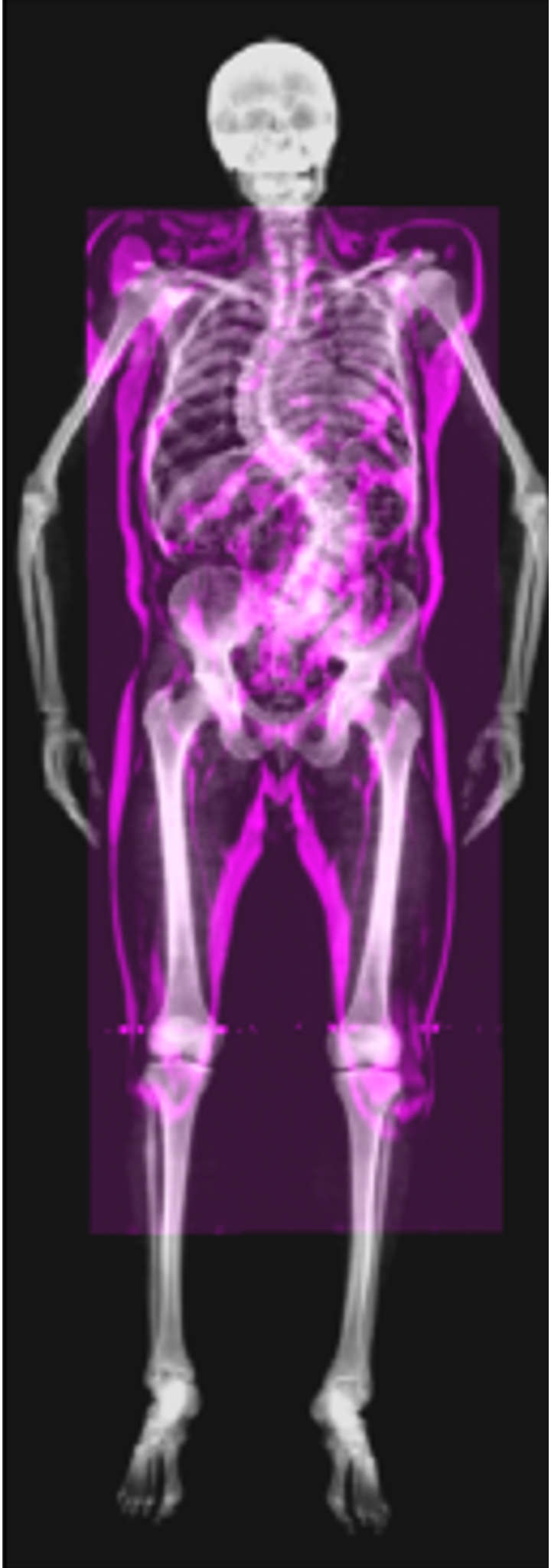}& 
            \includegraphics[trim= 0.0cm 1.7cm 0.0cm 0cm,clip, width=0.16\linewidth]{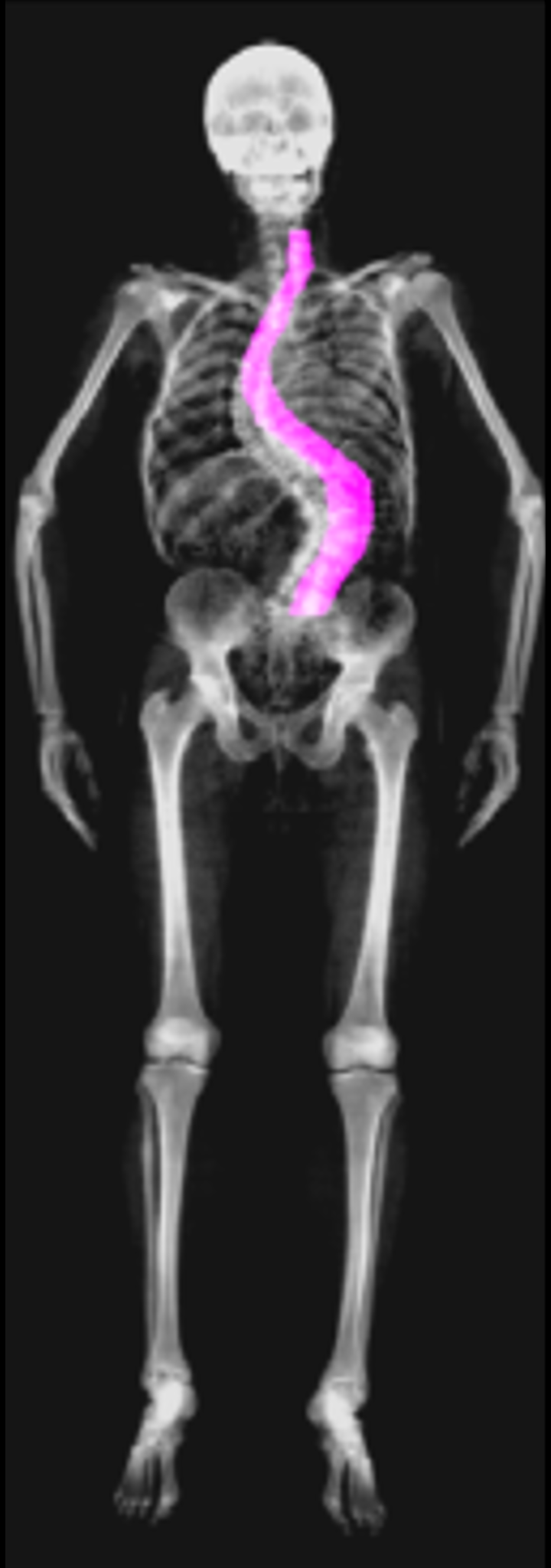}& 
            \includegraphics[trim= 0.0cm 1.7cm 0.0cm 0cm,clip, width=0.16\linewidth]{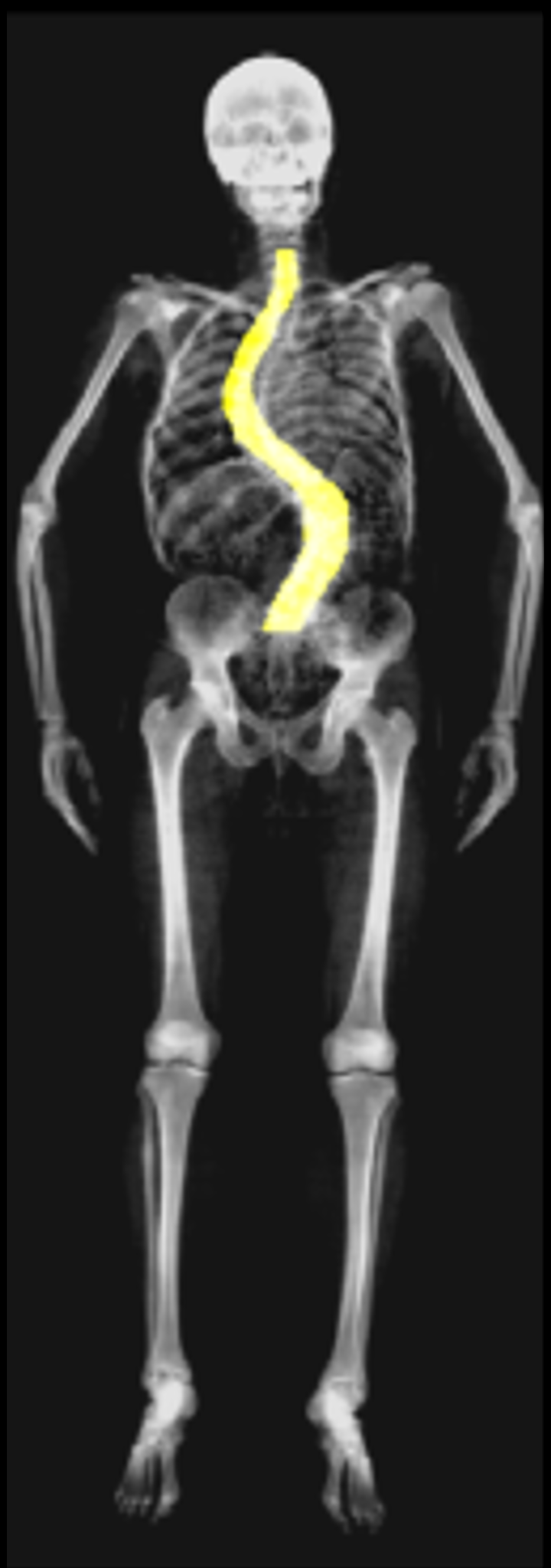}
        \end{tabular} } 
        \caption{\textbf{DXA to MRI two-Stage Alignment}. The two scans are iteratively aligned using a three parameter planar transformation. From left to right: DXA scan;  original coronal projection of MRI scan (not-aligned to the DXA); overlay of MRI aligned to DXA after the image-level alignment first stage; overlay of segmented spines after the first stage; overlay of spine segmentation after the  spine-level alignment second stage.}
    \label{fig:DXA_MRI_registration}
\end{figure}

\subsection{Problem definition}

The problem consists of regressing a 3D spine curve from a DXA image. This can be separated into two separate regression problems, namely: (i) the regression of the coronal or AP curve, and (ii) the regression of the sagittal or lateral curve.
For each of the 2D projections, we define three sets of points $P_{i} = \{ (x_i (z),y_i (z)), z\in\real \mid 1\le z\le 209\,\}, i={1,2,3}$ for the central and two lateral curves of the spine respectively, and $z$ is the vertical height of the scans (normalised between 1 and 209). These define the segmentation and mid-curve of the spine in the 2D projections. Our objective is to regress these curves (in the coronal and sagittal planes) from the DXA scan.

\begin{figure}[t]
\centering
\includegraphics[width=1\linewidth,height=3\linewidth,keepaspectratio]{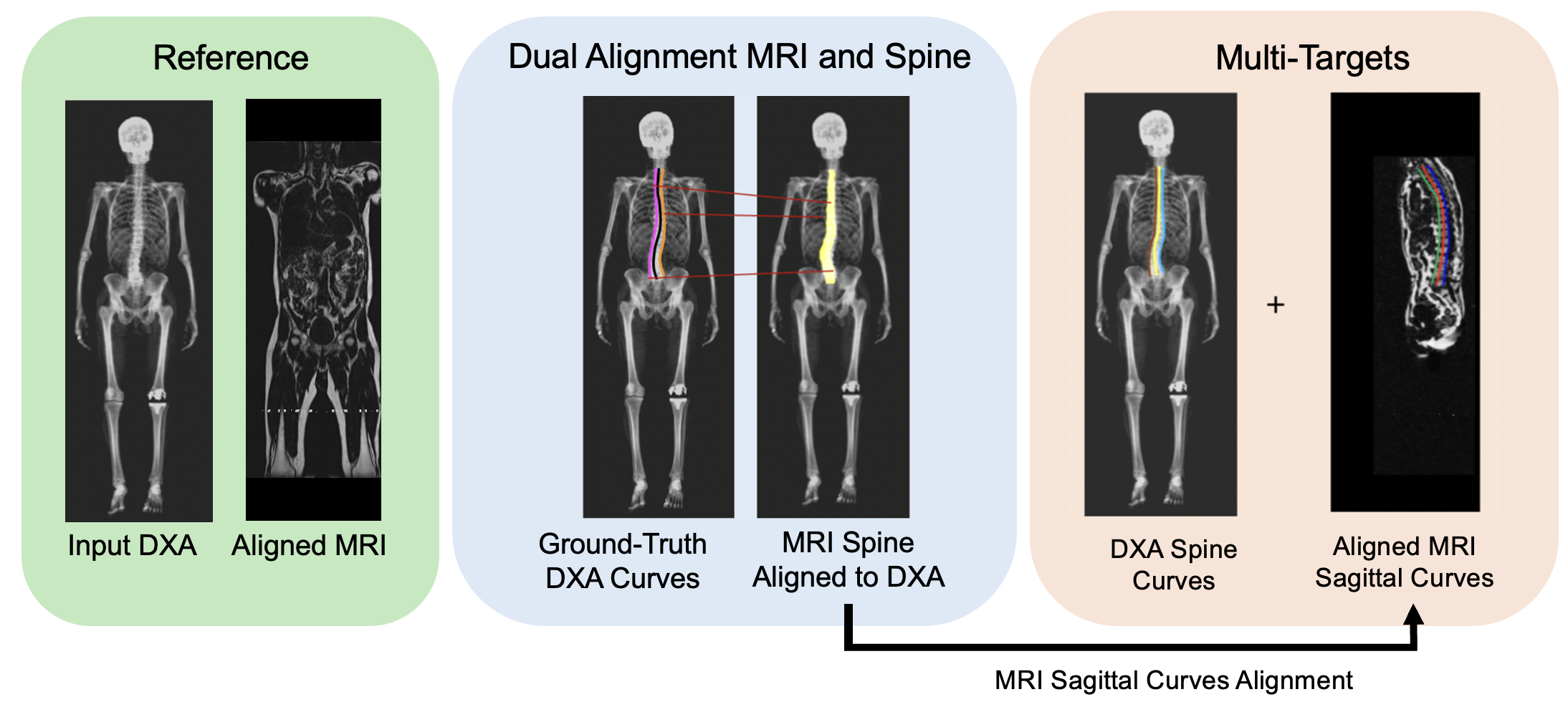} 
\caption{\textbf{Model Learning}. The regression model is learnt from pairs of aligned DXA and MRI scans. The regression targets are the DXA curve,  and the sagittal curve (projected from the 3D MRI spine). The alignment for the sagittal curve to DXA is obtained from the alignment of the coronal projection of the 3D MRI to the DXA. Six curves are regressed: the centerline of the spine as well as the left and right boundaries of the segmentation, for both the coronal and sagittal views.}

\label{fig:overview}
\end{figure}

\subsection{Modality Alignment}
\label{sec:alignment}

The paired images, the 2D DXA scans and the 3D MRIs, do not come registered. Hence, the first step we take is to align these two modalities. As we are interested in inferring the 3D information from 2D images; we register the 3D MRI and the extracted spine curve to the 2D DXA. The alignment proceeds iteratively in two stages: (i) a rough image-to-image alignment of the two imaging modalities followed by, (ii) a finer alignment of the extracted segmentation/curve of the spine from the MRI to the segmentation/curve of the spine from the DXA.
See Figure~\ref{fig:DXA_MRI_registration}.

For the rough alignment, we use a pipeline proposed by~\cite{Windsor21} which finds the best 2D transformation to align the 3D MRI (via its coronal projection) to the 2D DXA. 
The 2D transformations involve three parameters (a rotation angle $\theta$,  and two translations). 
To compute the transformation, 10 rotation angles are sampled in the
range $[-2,2]$ degrees, and the translation is obtained by convolving CNN spatial-feature maps of the DXA and MRI, and selecting the point of maximum response. 

For the second stage,  we align the spine curves between the two modalities. Again, the 2D transformation consists of a rotation and translation. We use keypoint matching sampled along the spine contour of the DXA, and  compute the transformation that minimises the mean squared error (MSE).   

It is possible that the person changed position too much between the DXA and MRI scans, and it is not possible to align the spines due to deformations. To check for this we measure the overlap of the spine segmented in the DXA scan with the projection of the 3D spine segmented in the MRI (see Figure~\ref{fig:DXA_MRI_registration}). We apply a threshold for filtering out the poorly aligned scans: if the Intersection-over-Union (IoU) of spine masks is below 70\%, then they are discarded. A total of 14,065 paired DXA-MRIs are rejected by this test which is 28.8\% of the data.

\subsection{Learning the Regressor}
\label{sec:regression}

In total, we regress 6 curves using a single model. The target curves for the regression are obtained from the DXA for the coronal view, and from the projection of the 3D MRI for the sagittal view (see Figure~\ref{fig:overview}).

\begin{figure}[t]
\includegraphics[width=\textwidth]{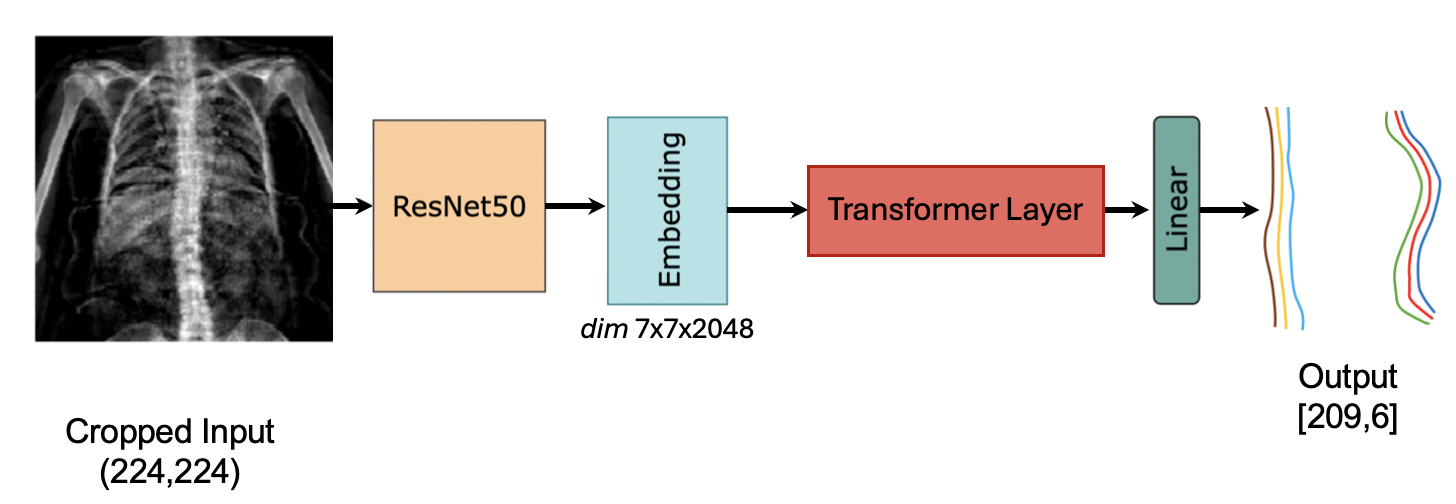}
\caption{\textbf{Image-Based Regression of Coronal and Sagittal Spine Curves.} We use a ResNet50,  pre-trained on ImageNet-21k, with a transformer layer to regress the spine curves  
$(x_{(1,2,3)}(z),y_{(1,2,3)}(z)), z \in [1,209]$ for left, center and right curves. The feature map extracted from ResNet50 are of resolution 7 x 7 x 2048, each vector feature from ResNet50 (49 x 2048) is used as input into a transformer layer. The model regresses the 6 curves (209 x 6) where we have 6 vectors for the 6 output spine curves, of dimension 209. Detailed Architecture in Appendix~\ref{Appendix:Implementation Details and Ablation}  Figure~\ref{fig_supp:Regression_Architecture}.} 
\label{fig:regression_centroids}
\end{figure}

\noindent \textbf{Encoder.}
Our regressor consists of a multi-scale feature extractor coupled with a transformer layer to learn the long-range dependencies more effectively (see Figure~\ref{fig:regression_centroids}). The initial part of the architecture consists of an image feature extractor with ResNet50 pre-trained on ImageNet-21k. The feature map of the penultimate convolutional block is extracted as $7 \times 7 \times 2048$ and fed into a standard transformer layer. The output of the transformer layer is then average pooled before the ultimate linear layer. The regression head consists of a linear layer to predict the output curve points from the transformer output vector.

\noindent \textbf{Loss.} Our loss is  the $L_1$ difference between the target and predicted spine curve points: $L = \sum_{i=1}^{n} \left| {x}^{i} - \widehat{x}^{i} \right|$, where $n$ is the number of samples,
$x^i$ are the ground-truth spine points, and $\widehat{x^i}$  the predicted spine points at a given $z^i$. 

We regress three curves for each view (coronal $(x)$ and sagittal $(y)$): the central curve of the spine as well as the left and right curves bounding the segmentation. The central and lateral curve points are ($x_{1}$,$x_{2}$,$x_{3}$,$y_{1}$,$y_{2}$,$y_{3}$), where 1 and 3 are either the right/left or anterior/posterior curves depending on the plane projection, and 2 is the central curve of the spine. 

\section{Dataset \& Implementation Details}
The UKBiobank is a publicly available dataset of 48,384 full-body Dixon MRI paired with DXA scans \cite{Sudlow2015UKBA}. The UK Biobank MRIs are resampled to be isotropic and cropped to a consistent resolution (700 $\times$ 224 $\times$ 224). The dataset is split into 80:10:10 for training (27,816), validation (3,477), and testing (3,477) after filtering the non-aligned scans by the dual alignment procedure explained in Section \ref{sec:alignment}.
Our training-validation-test is balanced in scoliosis cases, each containing 20\% of scoliosis cases. Our model takes as input a cropped DXA image of the whole spine. This is achieved using a $224 \times 224$ cropping window of the spine given spine segmentation from \cite{bourigault2022scoliosis}. This is done by computing the midpoint of the segment joining the endpoints of the spine from the segmentation mask and using it as the centre of the square window. The original DXA resolution is (832 $\times$ 320).

\noindent \textbf{Obtaining the 2D spine from the DXA images.}
We obtain the spine centroids and 2D spine mask segmentation in DXA which involves spine segmentation using pseudo-labelling in an active learning framework~\cite{Jamaludin2018PredictingSI,bourigault2022scoliosis}.This is used as the target for the coronal view regressor.

\noindent \textbf{Obtaining the 3D spine from the MRIs.}
We used a segmentation network to obtain the 3D whole spine curve in the MRI (3D centroids and segmentation)~\cite{Bourigault23} trained on adjacent axial slices (n-1,n,n+1) to limit the loss of depth information. These labels are used for training our sagittal regression model.

\subsection{Implementation Details}
The regressor uses a the Bottleneck Transformer~\cite{Srinivas2021BottleneckTF} together with a ResNet-50 image encoder. Details on the model architecture are given in the Appendix (see Section \ref{Appendix:Implementation Details and Ablation}).

For curve regression, all DXA inputs and target spine curves are normalised to fixed height for spine ranging from 1 to 209 pixels. We train our model for 500 epochs. We use ﬁve-fold cross-validation, where we repeat validation on 5 stratified folds. The batch size is set to 16, optimizer is Adam \cite{kingma2017adam} with $\beta=(0.9,0.999)$, and the learning rate is initially set to $1e^{-4}$ with decay every 200 epochs. We used one 32GB Tesla V100 GPU. To reduce overfitting, we employ two techniques, using dropout with a probability of $p=0.3$ and we employ a regularizer to the L1 loss. We also use different augmentation techniques with cropping, image contrast and random Gaussian noise in training.

\section{Results}
\subsection{Evaluation Metrics}
We measure the performance of our model using the mean absolute error $MAE=\frac{1}{n}\sum_{z=1}^n|y_z-\widehat{y_z}|$, and the mean of the relative error $RE = \frac{1}{n}\sum_{z=1}^n \frac{|y_z-\widehat{y_z}|}{\widehat{y_z}}$ between the predicted points $\widehat{y_z}$  and the ground truth $y_z, z \in [1,209]$. The mean of $RE$ is always between 0 and 1, the lower the relative error the better. To assess the accuracy of whole spine predicted masks obtained from the area between the lateral curves compared to project ground-truth masks obtained from segmentation of the MRI, we use the 2D Intersection-over-Union, $IoU$. We compute the 3D IoU to evaluate 3D spine shape estimation from our model given reference labels from MRIs, as explained in Section \ref{Appendix:3D Spine Curve from 2D}.

\subsection{Spine Curves Estimation and Robustness}
\label{sec:Spine Curves Estimation and Robustness}

\begin{figure}[t!]
    \centering
    \resizebox{1\linewidth}{!}{
    \tabcolsep=0.05cm

            \begin{tabular}{ccccccc}
            Input Coronal & GT Sagittal Curves & Predicted Sagittal Curves  & GT Sagittal Segmentation & Predicted Sagittal Segmentation & GT Spine & Predicted Spine \\ 
            \includegraphics[trim= 0.0cm 1.7cm 0.0cm 0cm,clip, width=0.16\linewidth]{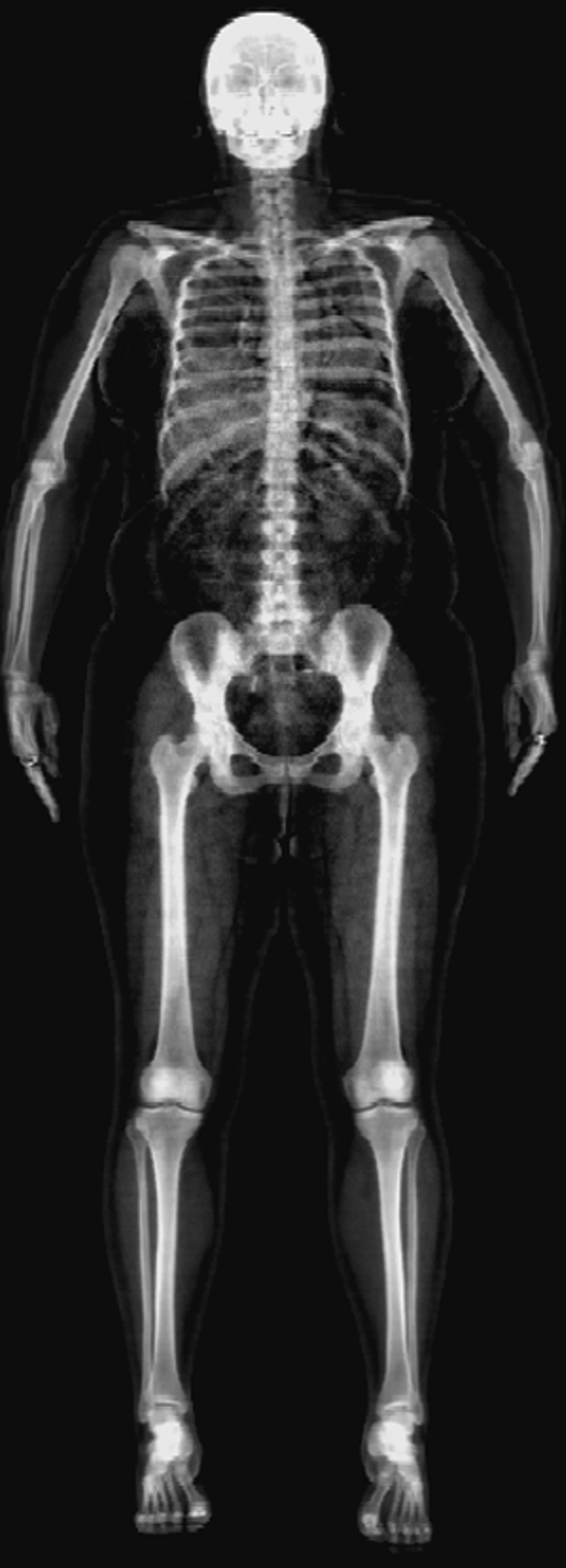}& 
            \includegraphics[trim= 0.0cm 1.7cm 0.0cm 0cm,clip, width=0.16\linewidth]{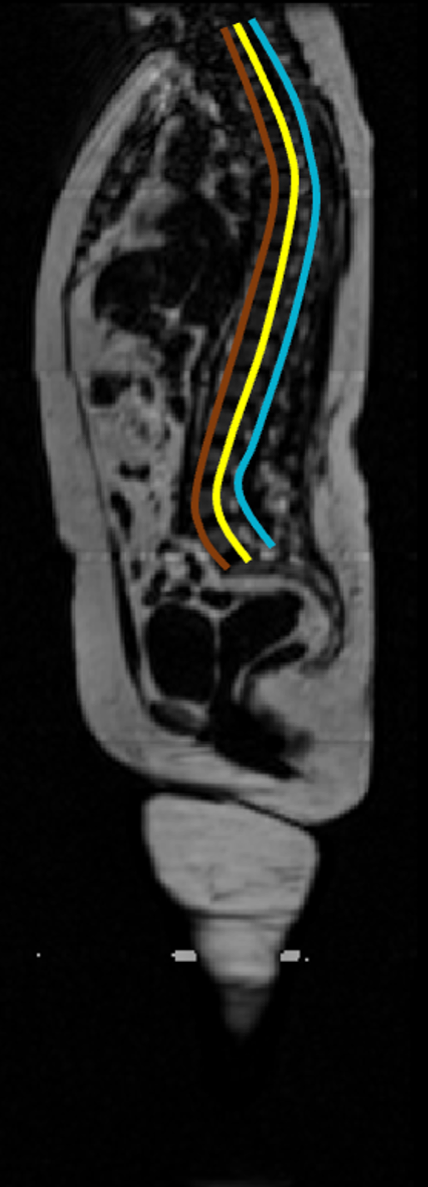}& 
            \includegraphics[trim= 0.0cm 1.7cm 0.0cm 0cm,clip, width=0.16\linewidth]{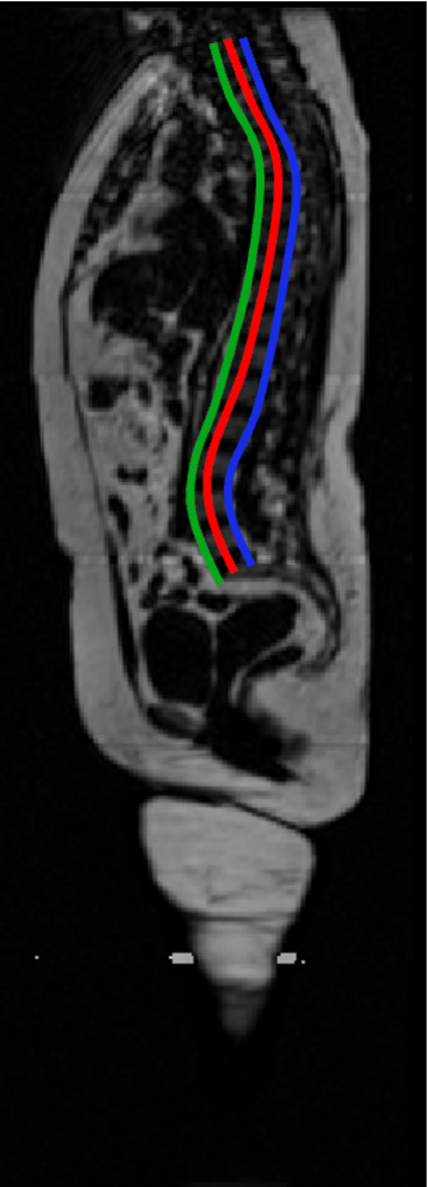}& 
            \includegraphics[trim= 0.0cm 1.7cm 0.0cm 0cm,clip, width=0.16\linewidth]{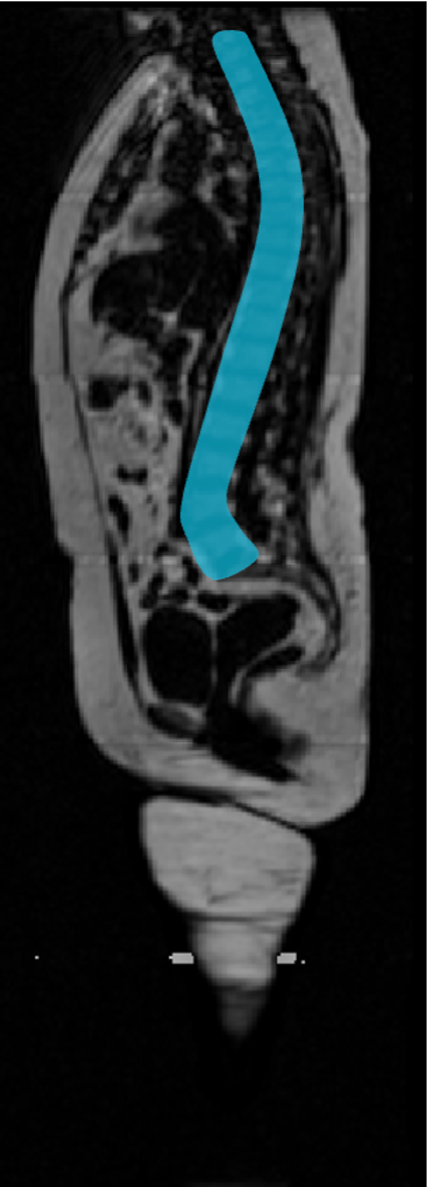}&
            \includegraphics[trim= 0.0cm 1.7cm 0.0cm 0cm,clip, width=0.16\linewidth]{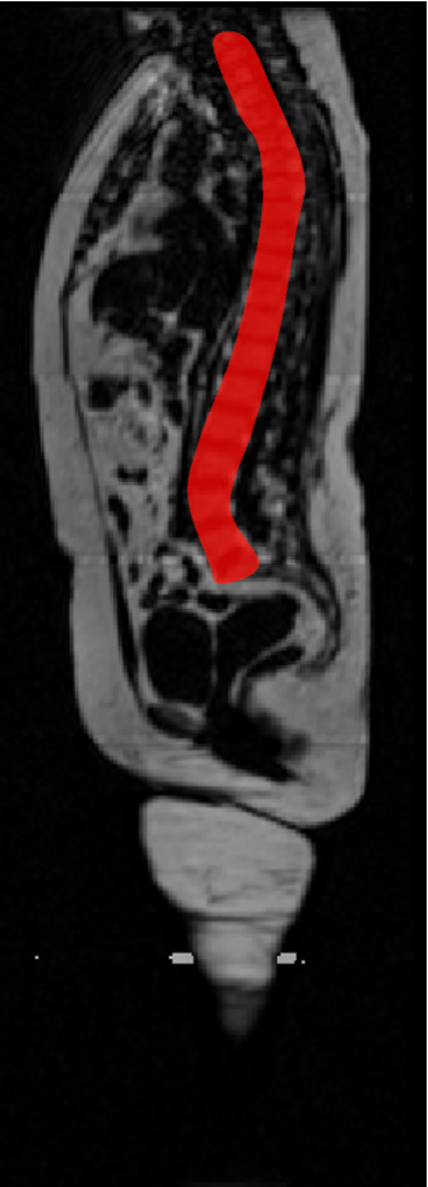}&
            \includegraphics[trim= 0.0cm 1.7cm 0.0cm 0cm,clip, width=0.16\linewidth]{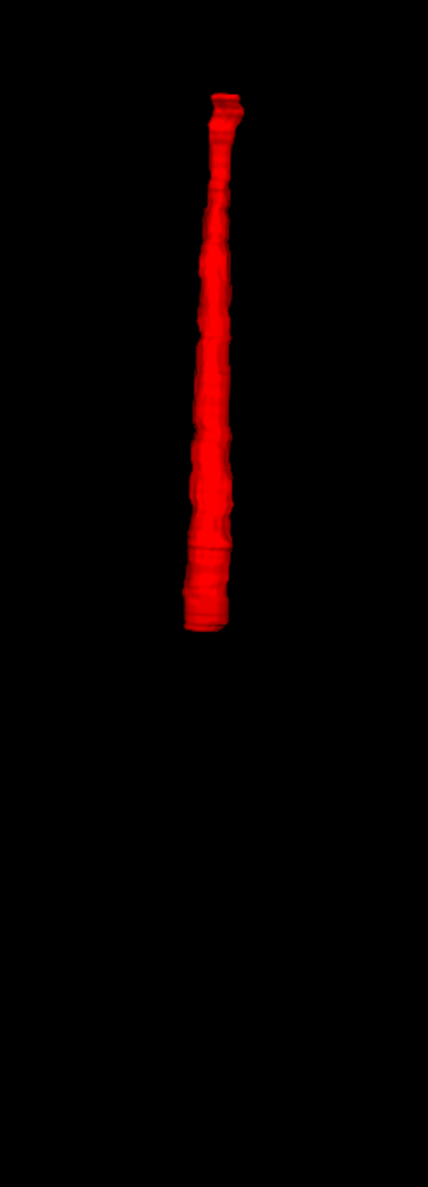}&
            \includegraphics[trim= 0.0cm 1.7cm 0.0cm 0cm,clip, width=0.16\linewidth]{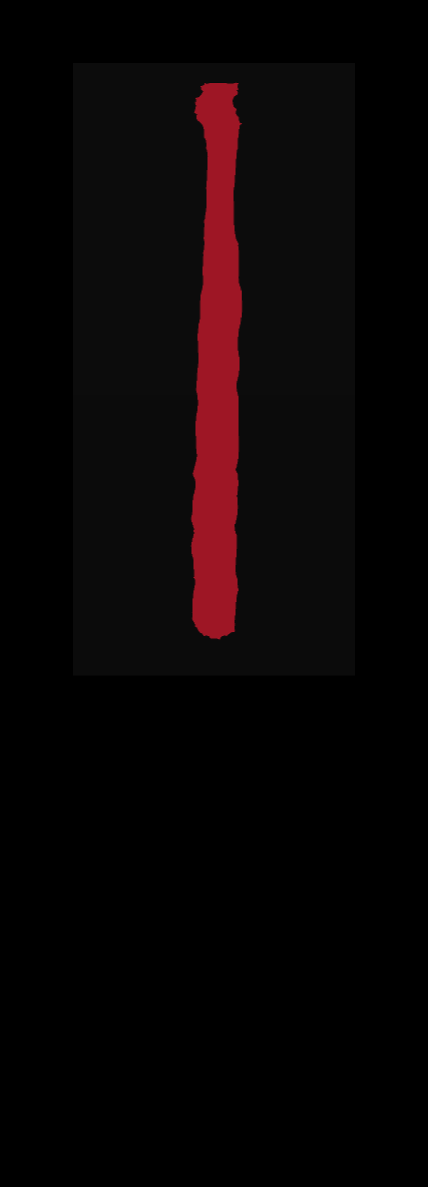}\\
            \hline 
            \\
            \includegraphics[trim= 0.0cm 1.7cm 0.0cm 0cm,clip, width=0.16\linewidth]{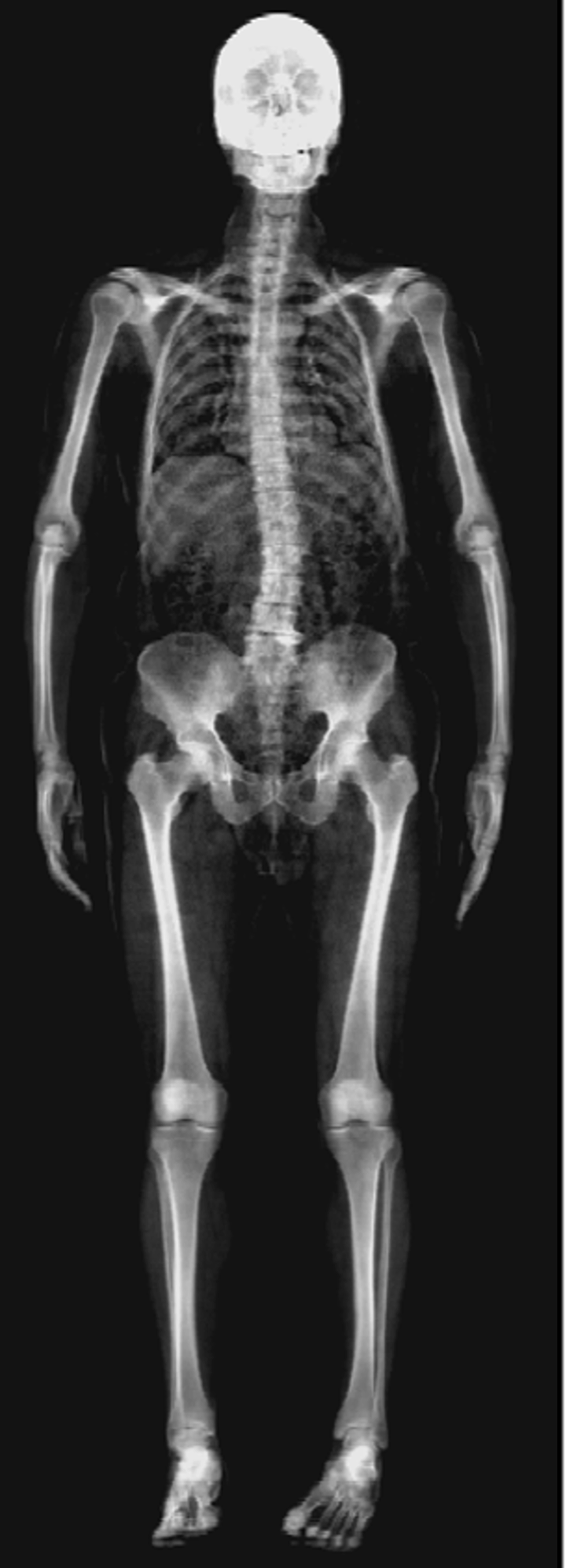}& 
            \includegraphics[trim= 0.0cm 1.7cm 0.0cm 0cm,clip, width=0.16\linewidth]{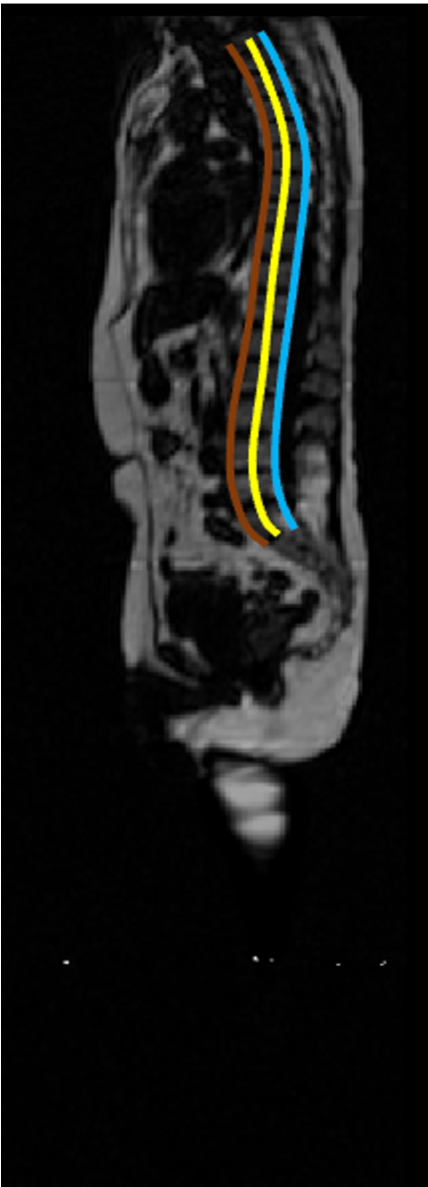}& 
            \includegraphics[trim= 0.0cm 1.7cm 0.0cm 0cm,clip, width=0.16\linewidth]{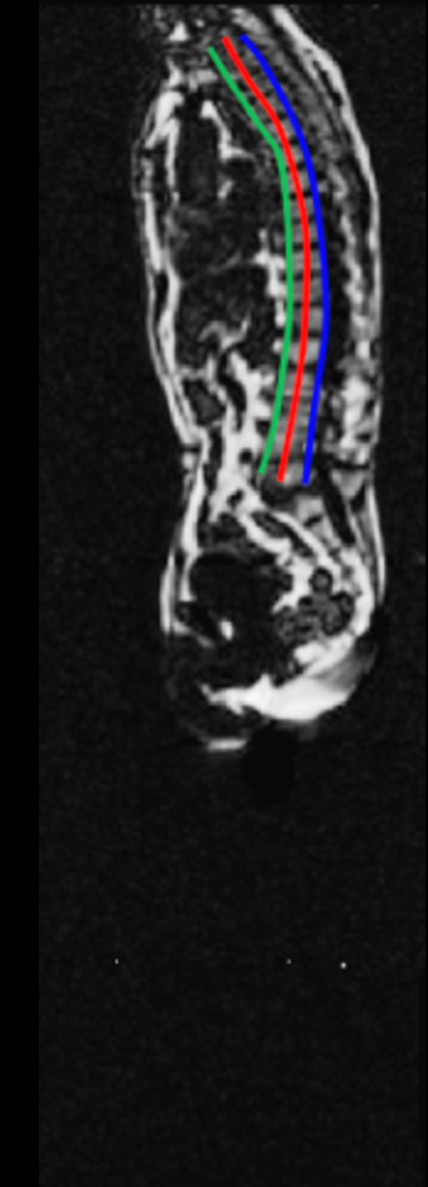}& 
            \includegraphics[trim= 0.0cm 1.7cm 0.0cm 0cm,clip, width=0.16\linewidth]{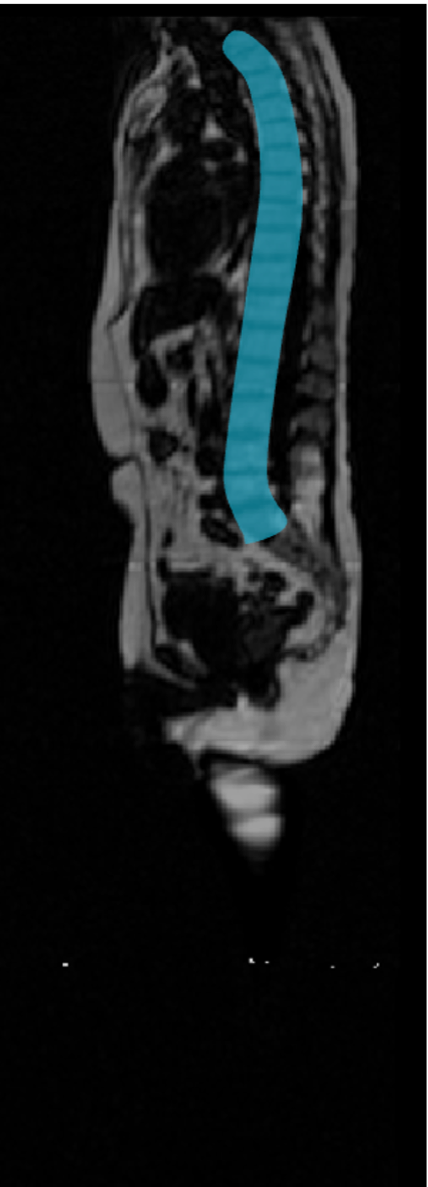}&
            \includegraphics[trim= 0.0cm 1.7cm 0.0cm 0cm,clip, width=0.16\linewidth]{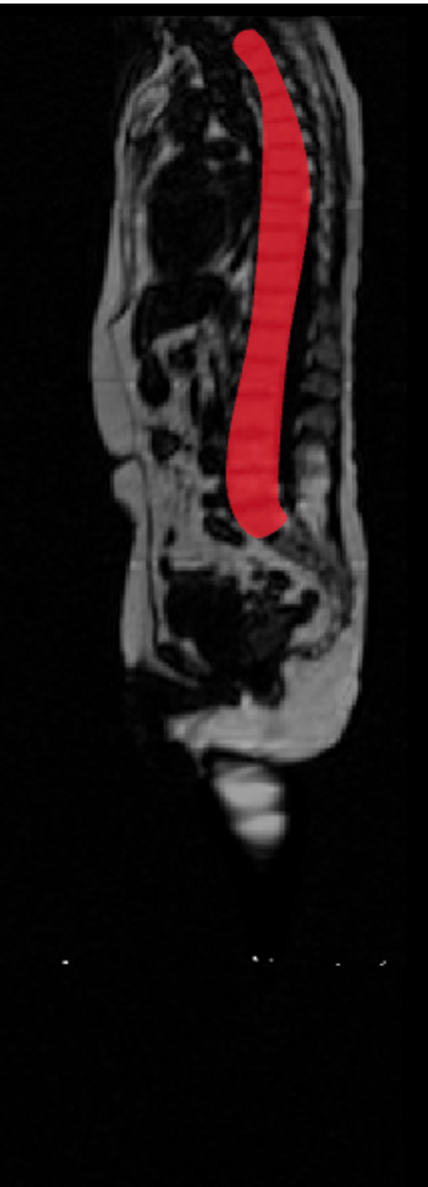}&
            \includegraphics[trim= 0.0cm 1.7cm 0.0cm 0cm,clip, width=0.16\linewidth]{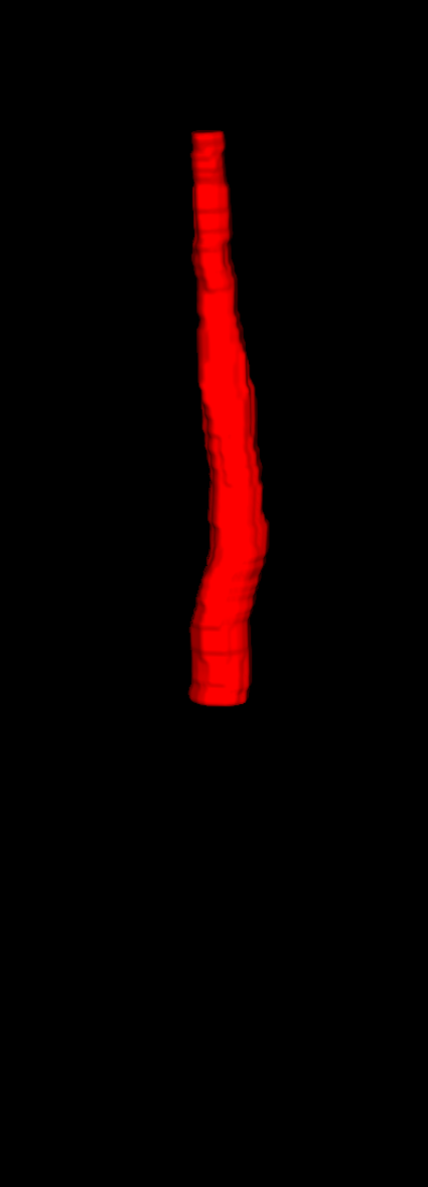}&
            \includegraphics[trim= 0.0cm 1.7cm 0.0cm 0cm,clip, width=0.16\linewidth]{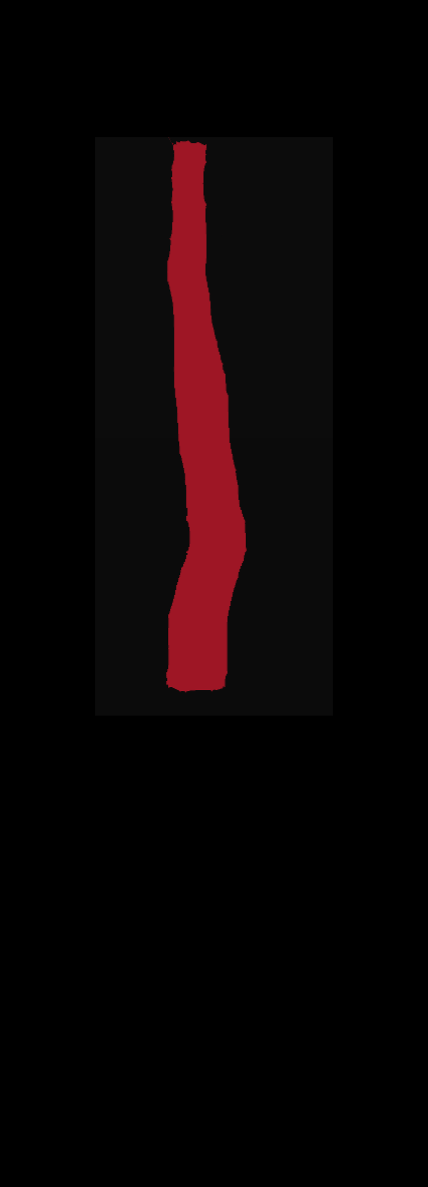}\\
            \hline 
            \\
            \includegraphics[trim= -0.12cm 1.7cm 0.0cm 0cm,clip, width=0.16\linewidth]{figs/Registration/Slide1.png}&
            \includegraphics[trim= 0.0cm 1.7cm 0.0cm 0cm,clip, width=0.16\linewidth]{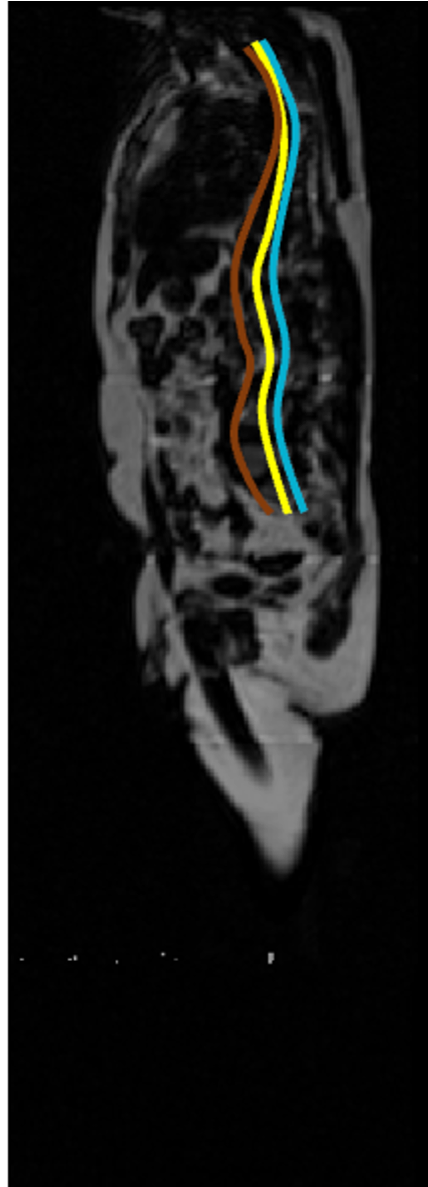}&
            \includegraphics[trim= 0.0cm 1.7cm 0.0cm 0cm,clip, width=0.16\linewidth]{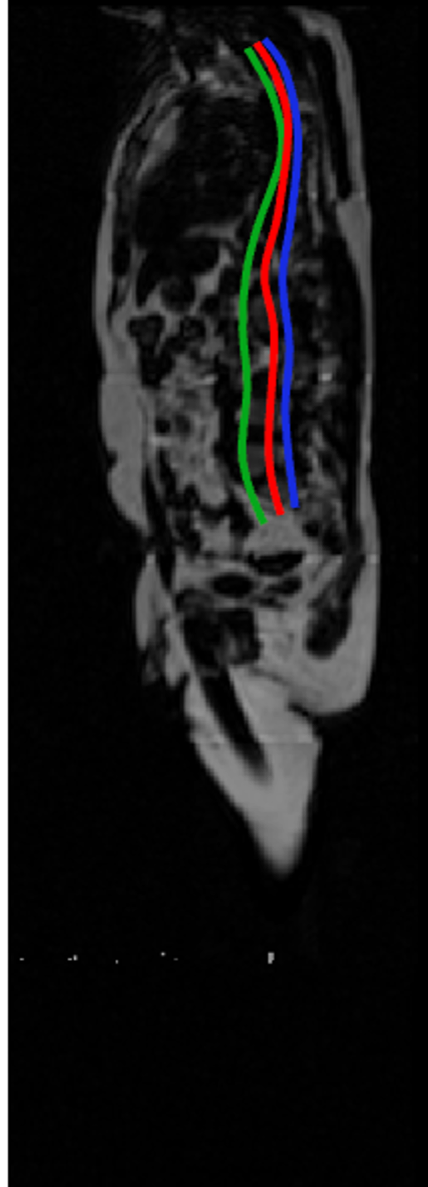}&
            \includegraphics[trim= 0.0cm 1.7cm 0.0cm 0cm,clip, width=0.16\linewidth]{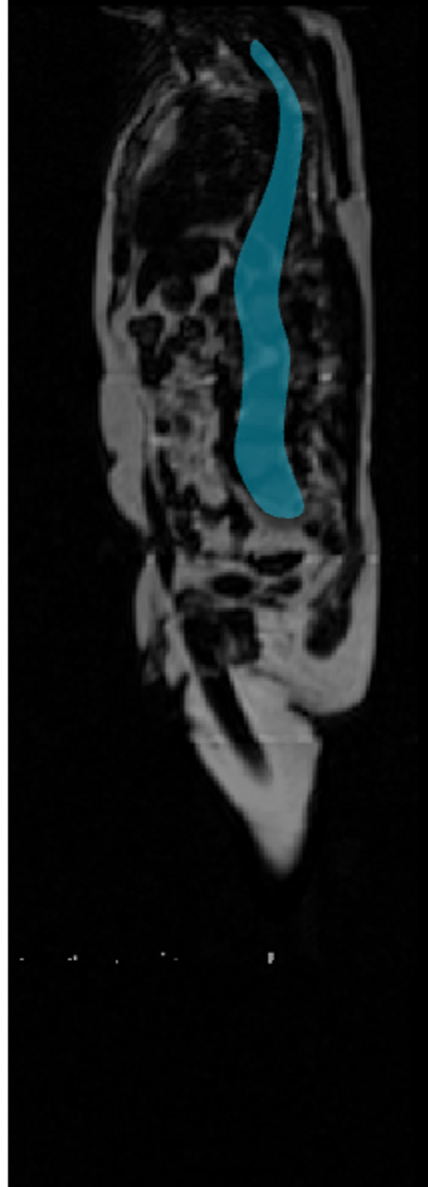}&
            \includegraphics[trim= 0.0cm 1.7cm 0.0cm 0cm,clip, width=0.16\linewidth]{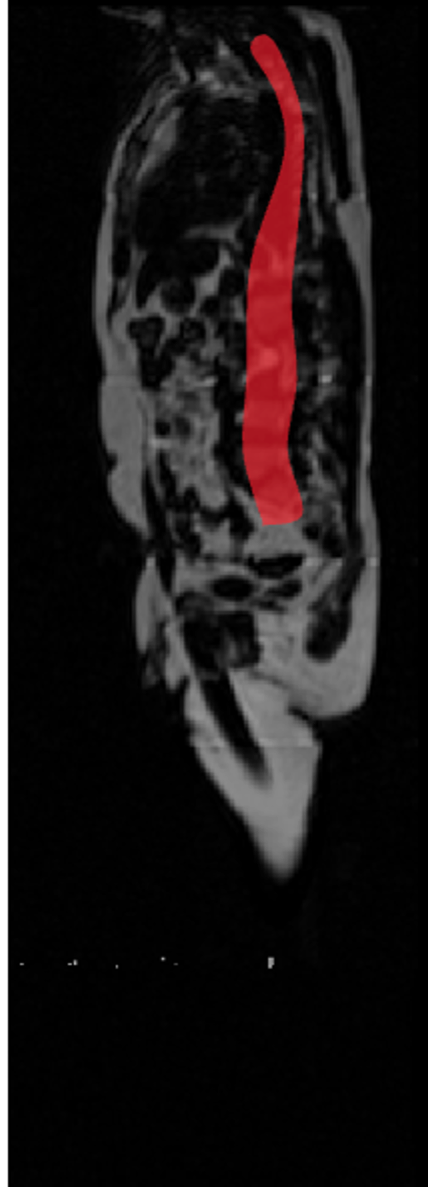}
            &\includegraphics[trim=-0.8cm 0cm 0.0cm 0cm,clip, width=0.16\linewidth]{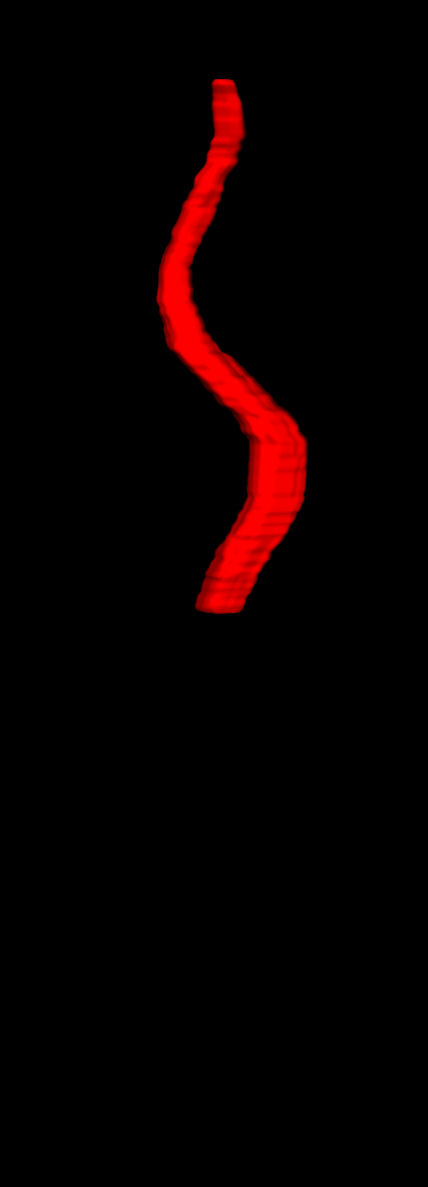}
            &\includegraphics[trim=-0.8cm 0cm 0.0cm 0cm,clip, width=0.16\linewidth]{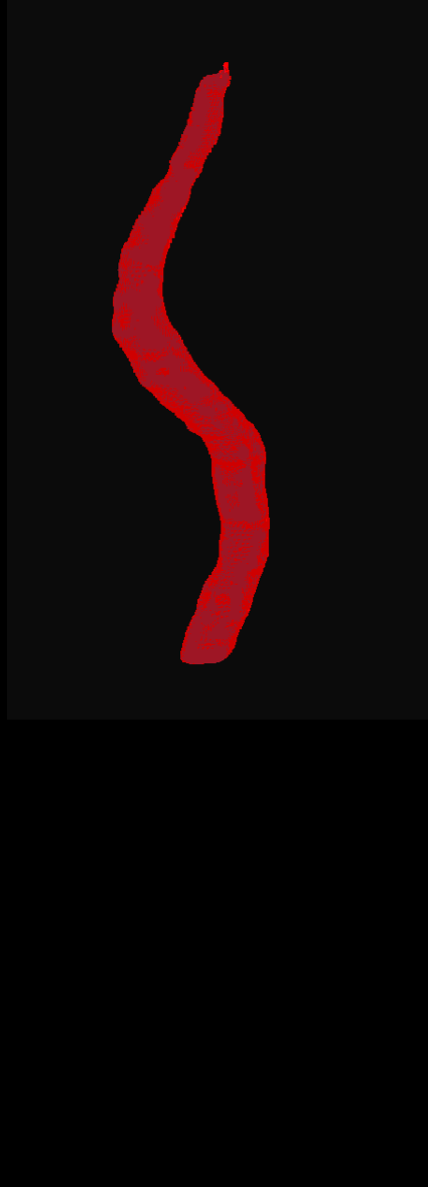}
        \end{tabular} } 
        \caption{\textbf{Qualitative Results of DXA to MRI Sagittal Curve Generation on the Test Set.} We show from left to right, the input DXA, the ground-truth sagittal curves, predicted sagittal curves, ground-truth spine mask, predicted spine mask overlayed on MRI sagittal slice (n=112), the ground-truth 3D spine, and its prediction using Gaussian rendering. The severity of the scoliosis increases from top to bottom.
        }
    \label{fig:Spine_Projections}
\end{figure}

The DXA and Coronal MRI spine curves are predicted at sub-pixel level precision with better precision and lower variance for the combined ResNet50+Transformer framework ($MAE=0.66 \pm0.21$, $RE=0.084 \pm0.02$, $IoU=89.6$) compared to ResNet50 or ViT (see Table~\ref{tb:spine_regression}).
We show that using residual blocks with a transformer layer, our model can reliably estimate not only the coronal MRI projection from a single DXA but any spine curve projections along a 360$^{\circ}$ rotation about the vertical axis such as the sagittal projections ($MAE=1.65 \pm0.6$, $RE=0.21 \pm0.08$, $IoU=86.8$), see Table~\ref{tb:spine_regression}). Our model is also able to capture the pattern of curves with straighter sagittal curves for more severe cases of scoliosis (see Figure~\ref{fig:Spine_Projections}).

\begin{figure}[t!]
\centering
 
\includegraphics[width=0.5\linewidth,height=0.5\linewidth,keepaspectratio]{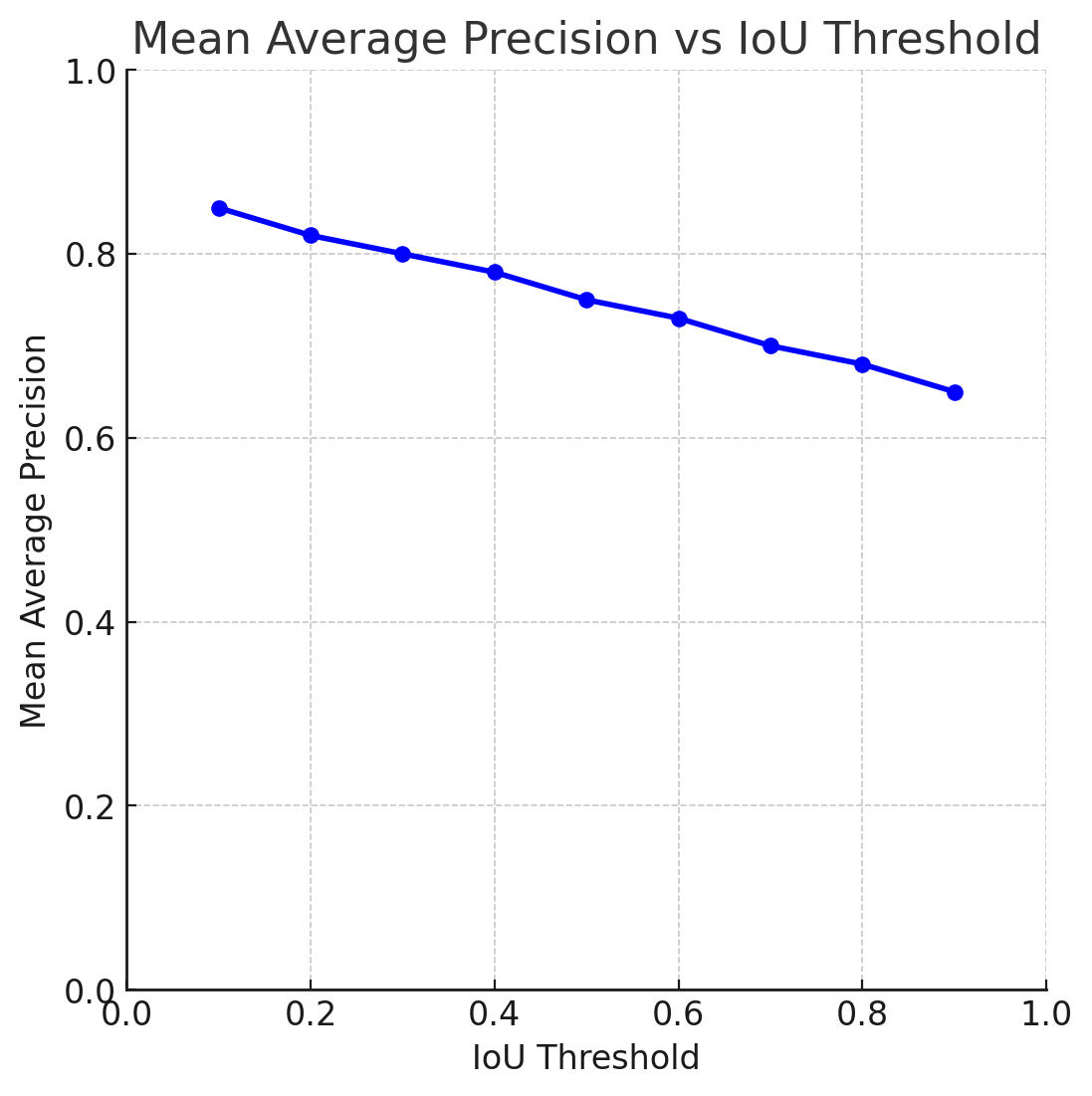} 
\caption{\textbf{Mean Average Precision of Predicted Spine Masks for Different IoU Thresholds over the Test samples.} We measure the performance of the 3D segmentation using 3D IoU. IoU thresholds ranges from 0.1 to 0.9, with steps of 0.1.}
\label{fig:mAP_IoU_thresholds}
\end{figure}

\noindent {\bf 3D evaluation.} We also evaluate the performance of our model to reconstruct 3D spines from 2D DXAs. We effectively have a segmentation of the spine bounded by the lateral curves from our pipeline. 
To obtain the 3D spine masks from two 2D segmentation, we use the 4 points in antero-posterior and left-right to generate a series of bounding axial plane ellipses as we go down in z (see Appendix~\ref{Appendix:3D Spine Curve from 2D}). We use an untouched set of 150 MRIs from the UKBiobank manually annotated spines from~\cite{Bourigault23} to measure the IoU.  The IoU between the predicted and ground-truth 3D spine masks, averaged over the test set, is $83.8 \pm 1.1$. We also measure the IoU for detected spine. See Figure~\ref{fig:mAP_IoU_thresholds} for visualising the mean average precision at different IoU thresholds.  We also measure the average deviation of the spine curve point-wise along the spine as a metric for 3D error. Our model achieves a 3D average spine curve deviation of $1.42$ voxels or $3.12mm$.
\noindent {\bf Discussion.} One probable reason that the model is able to infer a sagittal view from coronal DXA is that it has access to the entire (cropped) DXA scan, and there are at least two cues it can use: (i) the different intensities of the imaged bones of the spine give information about their angle and depth; and (ii) the position of the rib cage in the image depends on the spine, and so indirectly the 2D layout of the ribs in the image gives information about the spine.

\begin{table}[t!]
\centering
\resizebox{\textwidth}{!}{
\begin{tabular}{lccccccccl}
\hline
\multicolumn{1}{c}{\multirow{4}{*}{Target}} & \multirow{4}{*}{Model} & \multicolumn{3}{c}{\multirow{2}{*}{\begin{tabular}[c]{@{}c@{}}Spine Curves \\ Absolute Error\end{tabular}}} & \multicolumn{3}{c}{\multirow{2}{*}{\begin{tabular}[c]{@{}c@{}}Spine Curves \\ Relative Error\end{tabular}}} & \multicolumn{2}{c}{\multirow{2}{*}{Spine Mask (IoU)}} \\
\multicolumn{1}{c}{} & & \multicolumn{3}{c}{} & \multicolumn{3}{c}{} & \multicolumn{2}{c}{} \\ \cline{3-10} 
\multicolumn{1}{c}{} & & Mean & Median & SD & Mean & Median & SD & \multicolumn{2}{c}{} \\ \hline
& ResNet50 & 0.71 & 0.70 & $\pm$ 0.26 & 0.08 & 0.076 & $\pm$ 0.03 & \multicolumn{2}{c}{91.4} \\
DXA & ViT & 0.69 & 0.67 & $\pm$ 0.24 & 0.073 & 0.071 & $\pm$ 0.02 & \multicolumn{2}{c}{91.9} \\
& ResNet50 + Transformer & 0.58 & 0.57 & $\pm$ 0.18 & 0.062 & 0.057 & $\pm$ 0.01 & \multicolumn{2}{c}{92.3} \\ \hline
\multirow{3}{*}{Coronal MRI} & ResNet50 & 0.81 & 0.79 & $\pm$ 0.3 & 0.11 & 0.094 & $\pm$ 0.05 & \multicolumn{2}{c}{88.3} \\
& ViT & 0.78 & 0.78 & $\pm$ 0.28 & 0.091 & 0.083 & $\pm$ 0.03 & \multicolumn{2}{c}{88.9} \\
& ResNet50 + Transformer & 0.66 & 0.64 & $\pm$ 0.21 & 0.084 & 0.079 & $\pm$ 0.02 & \multicolumn{2}{c}{89.6} \\ \hline
\multirow{3}{*}{Sagittal MRI} & ResNet50 & 3.63 & 3.29 & $\pm$ 1.2 & 0.41 & 0.39 & $\pm$ 0.14 & \multicolumn{2}{c}{83.6} \\
& ViT & 2.99 & 2.73 & $\pm$ 1.1 & 0.38 & 0.36 & $\pm$ 0.11 & \multicolumn{2}{c}{84.1} \\
& ResNet50 + Transformer & 1.65 & 1.58 & $\pm$ 0.6 & 0.26 & 0.21 & $\pm$ 0.08 & \multicolumn{2}{c}{86.8} \\ \hline
\end{tabular}}
\vspace{2mm}
\caption{\textbf{Performance of Image-based Models for Spine Curves Regression.} 
For a given input DXA, we predict curves from the DXA target (\textit{1st block}) or curves from the corresponding MRIs (\textit{2nd and 3rd blocks}). We predict 3 curves (center + 2 laterals) for each block. The \textit{2nd and 3rd blocks} represent a single multi-view model outputting coronal and sagittal curves. The \textit{Target} column specifies the model configuration with output curve modality (DXA for baseline, coronal and sagittal MRI). The \textit{Model} column shows the different models. Then, we present the \textit{Absolute Error} and \textit{Relative Error} in terms of mean, median, and standard deviation of spine curves predicted versus reference curves (in pixels). The far-right column shows the 2D IoU of the masks bounded by the lateral spine curves.}
\label{tb:spine_regression}
\end{table}

\section{Conclusion}
Our model is able to give a patient-specific representation of the spine in 3D from a single DXA scan. We show that our method is effective in capturing the intricacies of the 3D spine. As such, this work has the potential to assist in the diagnosis and screening of scoliosis and other spinal disorders. Our primary focus for future work includes investigating confidence prediction for the sagittal curve.

\begin{credits}
\subsubsection{\ackname}
This work was supported by the Sustainable Approaches to Biomedical Science: Responsible and Reproducible Research (SABS: R³) Centre for Doctoral Training (EP/S024093/1), the EPSRC Programme Grant Visual AI (EP/T025872/1), and the Novartis-BDI Collaboration. This work has been conducted using the UK Biobank resource (application number 17295).

\subsubsection{\discintname}
The authors have no competing interests to declare that are relevant to the content of this article. 
\end{credits}

\bibliographystyle{splncs04}
\bibliography{egbib}

\appendix{}

\section{Implementation Details and Ablation}
\label{Appendix:Implementation Details and Ablation}

\begin{figure}[t]
\centering
\includegraphics[width=1\linewidth,height=3\linewidth,keepaspectratio]{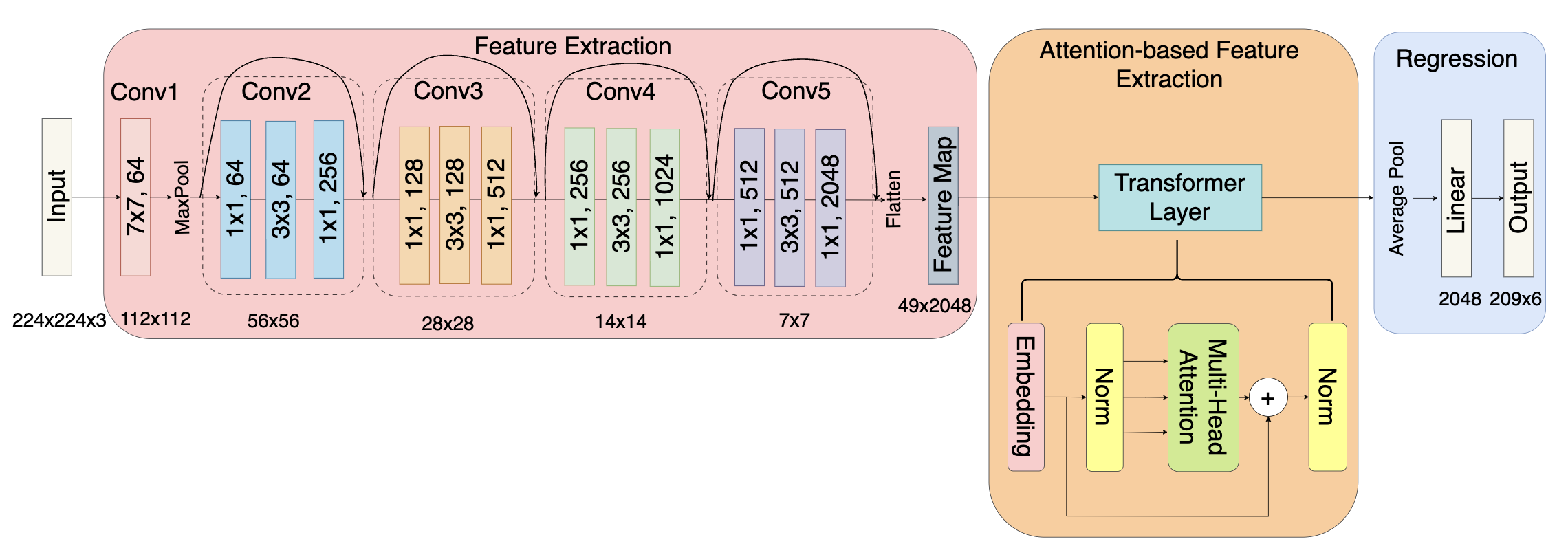} 
\caption{\textbf{Full Architecture of our ResNet50 with Transformer Layer.} 
}
\label{fig_supp:Regression_Architecture}
\end{figure}

\textbf{Regression Network of Spine Curves.}  
In this work we adopt a lightweight transformer layer on top of ResNet50 (see Figure~\ref{fig_supp:Regression_Architecture}). Before the final average pooling, we take the 7 x 7 x 2048 ResNet50 feature maps and flatten them to obtain 49 embedding tokens. These vectors are the input to the transformer layer. We follow the BotNet architecture \cite{Srinivas2021BottleneckTF} for the relative positioning encoding in the transformer layer. The output of the transformer layer is then average pooled before the ultimate linear layer. 



\noindent\textbf{Ablation Experiments.} We experiment with varying input sample size during training with an overall improved test performance for spine mask prediction of +3.4 IoU using the whole training set of 30k available versus 500 samples. Therefore, the network benefits from training on large datasets and it improves its ability to generalise. We also experiment adding more than one transformer layer on top of ResNet50. This does not significantly boost performance on the order of +0.002px and +0.003px average improvement for coronal and sagittal curve regression respectively.

\section{3D Spine from 2D Projections}
\label{Appendix:3D Spine Curve from 2D}

\begin{figure}[t]
\centering
\includegraphics[width=0.8\linewidth,height=3\linewidth,keepaspectratio]{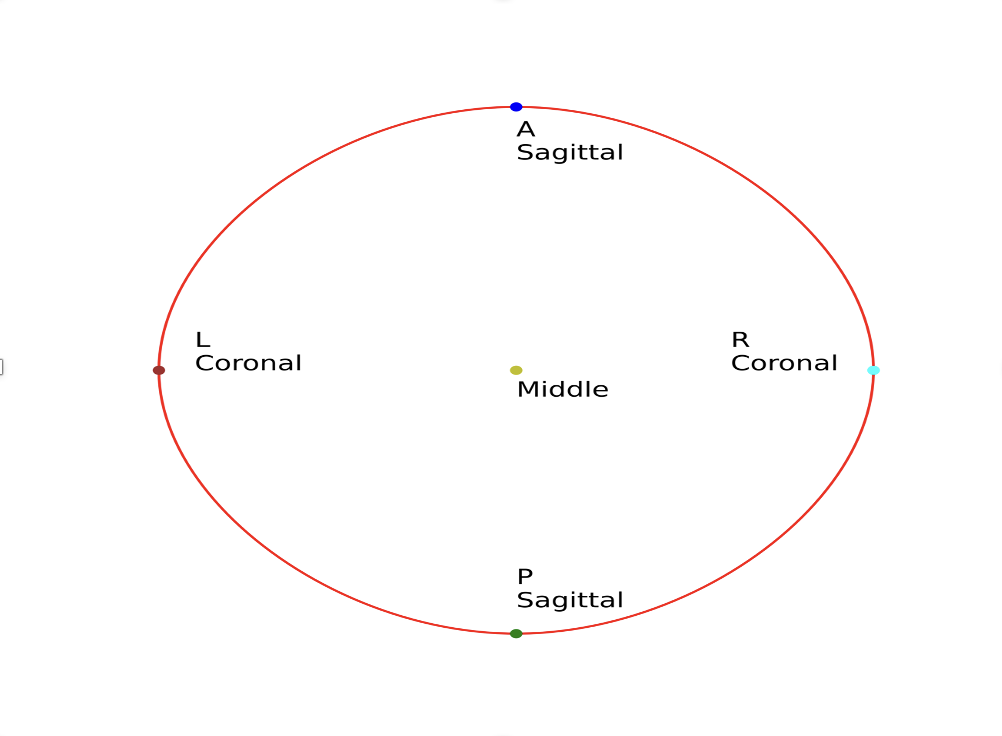} 
\caption{\textbf{3D Spine Shape Reconstruction through Ellipse Fitting.} We form ellipses in the axial plane along the spine top to bottom that go through key points from the orthogonal curves obtained by our model. We make sure the ellipses go through the antero-posterior (A-P) sagittal curves and the right-left (R-L) coronal curves. Each dot is a point on the 6 curves shown on Figure~\ref{fig:regression_centroids}. The middle point of the ellipse is made from the aligned mid coronal and mid sagittal curves.
}
\label{fig_supp:ellipse}
\end{figure}

\textbf{Computing 3D Spine Shape From 2D Orthogonal Curves.} In this section, we outline the steps taken for 3D spine shape recovery from two 2D planes i.e.\ coronal and sagittal. 
The output of our network are 2D spine curves on the coronal (XY) plane and sagittal (YZ) plane. 
We are able to reconstruct the 3D spine shape with minor post-processing. The idea is simple, fitting ellipses in the axial plane along the spine from top to bottom (see Figure~\ref{fig_supp:ellipse}). We ensure the ellipses go through predicted points from the two lateral coronal (right and left) and sagittal (antero and posterior) curves.

More examples of predictions of 3D spine shape from 2D DXA are available on our website \url{https://www.robots.ox.ac.uk/~vgg/research/dxa-to-3d}.
Our model works well in estimating sagittal MRI projections for normal spines and severe scoliosis spines.

\end{document}